\pdfoutput=1
\RequirePackage{ifpdf}
\ifpdf 
\documentclass[pdftex]{sigma}
\else
\documentclass{sigma}
\fi

\begin{document}

\allowdisplaybreaks

\renewcommand{\thefootnote}{$\star$}

\renewcommand{\PaperNumber}{047}

\FirstPageHeading

\ShortArticleName{$\mathcal{PT}$ Symmetry and QCD: Finite Temperature and Density}

\ArticleName{$\boldsymbol{\mathcal{PT}}$ Symmetry and QCD:\\ Finite Temperature and Density\footnote{This paper is a contribution to the Proceedings of the VIIth Workshop ``Quantum Physics with Non-Hermitian Operators''
     (June 29 -- July 11, 2008, Benasque, Spain). The full collection
is available at
\href{http://www.emis.de/journals/SIGMA/PHHQP2008.html}{http://www.emis.de/journals/SIGMA/PHHQP2008.html}}}

\Author{Michael C. OGILVIE  and Peter N. MEISINGER}

\AuthorNameForHeading{M.C. Ogilvie and P.N. Meisinger}

\Address{Department of Physics, Washington University, St. Louis, MO 63130, USA}
\Email{\href{mailto:mco@physics.wustl.edu}{mco@physics.wustl.edu}, \href{mailto:pnm@physics.wustl.edu}{pnm@physics.wustl.edu}}

\ArticleDates{Received November 15, 2008, in f\/inal form April 10,
2009; Published online April 17, 2009}

\Abstract{The relevance of $\mathcal{PT}$ symmetry to quantum chromodynamics (QCD), the
gauge theory of the strong interactions, is explored in the context
of f\/inite temperature and density. Two signif\/icant problems in QCD
are studied: the sign problem of f\/inite-density QCD, and the problem
of conf\/inement. It is proven that the ef\/fective action for heavy
quarks at f\/inite density is $\mathcal{PT}$-symmetric. For the case of $1+1$
dimensions, the $\mathcal{PT}$-symmetric Hamiltonian, although not Hermitian,
has real eigenvalues for a range of values of the chemical potential~$\mu$, solving the sign problem for this model.
The ef\/fective action for heavy quarks is part of a~potentially large class
of generalized sine-Gordon models which are non-Hermitian
but are $\mathcal{PT}$-symmetric.
Generalized sine-Gordon models also occur naturally
in gauge theories in which magnetic monopoles lead
to conf\/inement. We explore gauge theories
where monopoles cause conf\/inement at arbitrarily high temperatures.
Several dif\/ferent classes of monopole gases exist, with each class leading
to dif\/ferent string tension scaling laws.
For one class of monopole gas models, the $\mathcal{PT}$-symmetric af\/f\/ine Toda f\/ield theory
emerges naturally as the ef\/fective theory. This in turn leads to
sine-law scaling for string tensions,
a behavior consistent with lattice simulations.}


\Classification{81T13; 81R05; 82B10}

\section[$PT$ symmetry and two difficult problems of QCD]{$\boldsymbol{\mathcal{PT}}$ symmetry and two dif\/f\/icult problems of QCD}

Models with $\mathcal{PT}$ symmetry have emerged as an interesting extension
of conventional quantum mechanics. There is a large class of models
that are not Hermitian, but nevertheless have real spectra as a consequence
of $\mathcal{PT}$ symmetry. Bender and Boettcher have shown that single-component
quantum mechanical models with $\mathcal{PT}$-symmetric potentials of the form
$-\lambda\left(-ix\right)^{p}$ have real spectra \cite{Bender:1998ke}.
An extensive literature on $\mathcal{PT}$ symmetry and related matters now
exists, and there are extensive review articles available \cite{Bender:2005tb,Bender:2007nj}.
Here we explore the relevance of $\mathcal{PT}$ symmetry for two of the
most dif\/f\/icult problems in quantum chromodynamics (QCD), the gauge
theory of the strong interaction.

The sign problem of QCD arises in the Euclidean space approach to
QCD at f\/inite, i.e., non-zero, quark number density
\cite{Stephanov:2007fk,Lombardo:2008sc}.
There
is broad interest, both theoretically and experimentally, in the properties
of QCD at f\/inite temperature and density.
Finite density QCD is particularly important for exploring
the possibility of color-superconducting
quark matter in the interiors of neutron stars \cite{Alford:2007xm}. Lattice gauge
theory has proven to be a powerful tool for exploring QCD and related
models at f\/inite temperature.
Unfortunately, these results
have been obtained largely for zero density. Non-zero quark density
is implemented by introducing a chemical potential $\mu$ for quark
number. Within the Euclidean space formalism,
a non-zero temperature $T$ is obtained by making the bosonic f\/ields
periodic in Euclidean time, with period $\beta = 1/T$. This
is easy to implement in lattice simulations.
Non-zero chemical potential, on the other hand, must be implemented
in a way that makes the weight function used in the
Feynman path integral complex. This is the so-called sign problem
of f\/inite density QCD. The complex weight assigned to Euclidean f\/ield
conf\/igurations spoils the probabilistic interpretation of the Euclidean
path integral, making the use of conventional importance-sampling
algorithms impossible. While there have been impressive ef\/forts to
simulate f\/inite-density QCD by extrapolating from $\mu=0$, the sign
problem remains a dif\/f\/icult, fundamental, and important problem.
We will show below that QCD at f\/inite density may be interpreted
as a theory with $\mathcal{PT}$ symmetry. We will show explicitly
how a $(1+1)$-dimensional gauge model can be reduced to
a $\mathcal{PT}$-symmetric Hamiltonian over the gauge group,
with real eigenvalues for a range of values of $\beta\mu$.

\begin{figure}[t]
\centerline{\includegraphics[width=8cm]{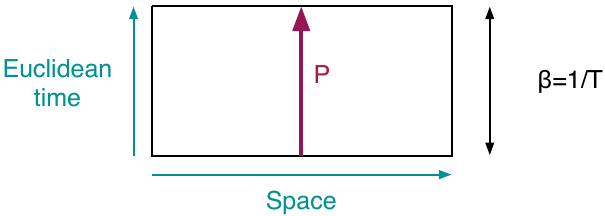}}
\caption{The Polyakov loop in Euclidean space-time.}\label{Fig1}
\end{figure}

The other problem of modern strong-interaction physics
we will consider
is the origin of quark conf\/inement.
In many ways, it is the most important problem
in QCD, because the conf\/inement of quarks
inside hadrons is the fundamental
property of QCD not fully understood
theoretically.
A good overview of various approaches
to this problem is provided by
the review of Greensite \cite{Greensite:2003bk}.
Finite temperature gauge theories
are advantageous in many aspects for the study of conf\/inement.
This is due largely to the utility and ubiquity of the Polyakov loop operator.
Def\/ined as a path-ordered exponential of the gauge f\/ield,
in $3+1$ dimensions the
Polyakov loop operator $P$ is given by
\begin{gather*}
 P\left(\vec{x}\right)=\mathcal{P}\exp\left[i\int_{0}^{\beta}dtA_{4}\left(\vec{x},t\right)\right],
\end{gather*}
and represents the insertion of a static quark into a thermal system
of gauge f\/ields at a temperature $T=\beta^{-1}$.
Fig.~\ref{Fig1} shows the Polyakov loop in this geometry.
Because of the periodic boundary conditions in the Euclidean time direction,
the Polyakov loop is a closed loop, and its trace
is gauge invariant.
Also known as the Wilson line, the Polyakov loop
represents the insertion of a static quark at a spatial point $\vec{x}$
in a gauge theory at f\/inite temperature. In particular,
the thermal average of the trace of $P$ in an
irreducible representation $R$ of the gauge group
is associated with  the additional free energy $F_{R}$ required to insert a static
quark in the fundamental representation via
\begin{gather*}
\langle  {\rm Tr}_{R}P\left(\vec{x}\right)\rangle  =e^{-\beta F_{R}}.
\end{gather*}
Pure ${\rm SU}(N)$ gauge theories have a global $Z(N)$ symmetry $P\rightarrow zP$
where $z=e^{\frac{2\pi i}{N}}$ is the generator of $Z(N)$, the center
of ${\rm SU}(N)$. This symmetry, if unbroken, guarantees that for the fundamental
representation $F$,
$\langle  {\rm Tr}_{F}P\left(\vec{x}\right)\rangle  =0$.
This is interpreted as $F_{F}$ being inf\/inite, and an inf\/inite free
energy is required to insert a heavy quark into the system. On the
other hand, if the $Z(N)$ symmetry  is spontaneously broken, the free energy
required is f\/inite. Thus conf\/inement in pure gauge theories is associated
with unbroken center symmetry, and broken symmetry with a~deconf\/ined
phase. The Polyakov loop is the order parameter for the deconf\/inement
transition in pure gauge theories $\langle {\rm Tr}_{F}P\rangle =0$ in the
conf\/ined phase and $\langle {\rm Tr}_{F}P\rangle \ne0$ in the deconf\/ined phase.
The addition of dynamical quarks in the fundamental representation
explicitly breaks this $Z(N)$ symmetry. Nevertheless, the Polyakov
loop remains important in describing the behavior of the system,
as we will see in our treatment of the sign problem.

In pure gauge theories, the Wilson loop operator is used to
measure the string tension between quarks
in the conf\/ined phase where $F_R$ vanishes
for representations transforming non-trivially
under $Z(N)$.
At non-zero temperature,
a timelike string tension $\sigma_{k}^{(t)}$ between $k$ quarks and
$k$~antiquarks can be measured from the behavior of the correlation
function \begin{gather*}
\langle {\rm Tr}_{F}P^{k}\left(\vec{x}\right){\rm Tr}_{F}\left(P^{+}\left(\vec{y}\right)\right)^{k}\rangle \simeq\exp\left[-\frac{\sigma_{k}^{(t)}}{T}\left|\vec{x}-\vec{y}\right|\right]
\end{gather*}
at suf\/f\/iciently large distances.
A conf\/ining phase is def\/ined by two properties:
the expectation value $\langle {\rm Tr}_{R}P\rangle $
is zero for all representations $R$ transforming
non-trivially under $Z(N)$, and the
string tensions $\sigma_{k}^{(t)}$ must be non-zero
for $k=1$ to $N-1$.
There are two kinds of model f\/ield theories,
related to QCD, for which
these two properties are known to hold.
As we discuss below, $\mathcal{PT}$ symmetry plays
an interesting role, which may extend to QCD.

\section{The chemical potential and the sign problem}

Perturbation theory can be
used to calculate the one-loop free energy density
$f_{q}$ of quarks in $d+1$ dimensions
 in the fundamental representation with spin degeneracy
$s$ moving in a Polyakov loop background at non-zero temperature $T=\beta^{-1}$ and
chemical potential $\mu$
 \begin{gather*}
f_{q}=-sT\int\frac{d^{d}k}{\left(2\pi\right)^{d}}\, {\rm Tr}_{R}
\left[\ln\big(1+Pe^{\beta\mu-\beta\omega_{k}}\big)
+\ln\big(1+P^{+}e^{-\beta\mu-\beta\omega_{k}}\big)\right],
\end{gather*}
 where $\omega_{k}=\sqrt{k^2+M^2}$ is the energy of the particle as a function of
$k$ and $M$ is the mass of the particle
 \cite{Gross:1980br,Weiss:1980rj}.
The expression for a bosonic f\/ield is similar.
The logarithm can be expanded to give
\begin{gather*}
f_{q}=sT\int\frac{d^{d}k}{\left(2\pi\right)^{d}}
\sum_{n=1}^{\infty}\frac{\left(-1\right)^{n}}{n}\left[e^{n\beta\mu-n\beta\omega_{k}}
{\rm Tr}_{R}P^{n}+e^{-n\beta\mu-n\beta\omega_{k}}{\rm Tr}_{R}P^{+n}\right].
\end{gather*}
This expression has a simple interpretation as a sum of paths winding
around the timelike direction. With standard boundary conditions,
which are periodic for bosons and antiperiodic for fermions, this
one-loop free energy always favors the deconf\/ined phase.

The ef\/fects of heavy quarks in the fundamental representation,
 with $\beta M \gg 1$, on the
gauge theory can be obtained approximately from the $n=1$ term in
the free energy
\begin{gather*}
f_{q}\approx-sT\int\frac{d^{d}k}{\left(2\pi\right)^{d}}{\rm Tr}_{F}
\left[Pe^{\beta\mu-\beta\omega_{k}}+P^{+}e^{-\beta\mu-\beta\omega_{k}}\right],
\end{gather*}
 because term with higher $n$ are suppressed by a factor $e^{-n\beta M}$. In this approximation, bosons
and fermions have the same ef\/fect at leading order. After integrating
over $k$, the free energy $f_{q}$ can be written as $f_{q}\approx-h_{F}\left[e^{\beta\mu}{\rm Tr}_{F}P+e^{-\beta\mu}{\rm Tr}_{F}P^{+}\right]$.
The one-loop free energy density is the one-loop ef\/fective potential
at f\/inite temperature.
Thus the free energy for the heavy quarks can be added to the usual
gauge action to give an ef\/fective action which involves only
the gauge f\/ields.
The ef\/fective action is
given by
\begin{gather*}
S_{\rm ef\/f}=\int d^{d+1}x\left[\frac{1}{4g^{2}}\left(F_{\mu\nu}^{a}\right)^{2}
-h_{F}\big(e^{\beta\mu}{\rm Tr}_{F}(P)+e^{-\beta\mu}{\rm Tr}_{F}(P^{+})\big)\right]
\end{gather*}
and the structure and symmetries of the theory are obviously the same
in any number of spatial dimensions.
Because ${\rm Tr}_{F}P$ is complex for $N\ge3$, the ef\/fective
action for the gauge f\/ields is complex. This is a form of the so-called
sign problem for gauge theories at f\/inite density: the Euclidean path
integral involve complex weights. This problem is a fundamental
barrier to lattice simulations of QCD at f\/inite density.

\section[Heavy quarks at $\mu\ne0$ in $1+1$ dimensions]{Heavy quarks at $\boldsymbol{\mu \ne0}$ in $\boldsymbol{1+1}$ dimensions}

\begin{figure}[t]
\centerline{\includegraphics[width=8cm]{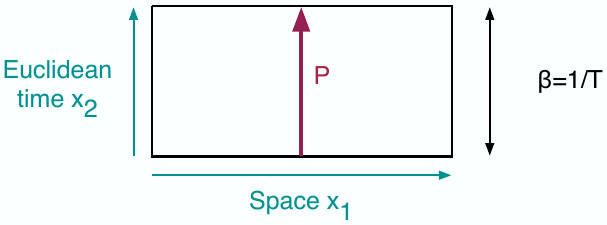}}
\caption{The Polyakov loop in $(1+1)$-dimensional space-time.}\label{Fig2}
\end{figure}

\begin{figure}[t]
\centerline{\includegraphics[width=3.5cm]{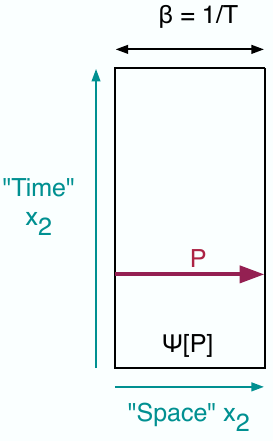}}
\caption{The Polyakov loop in a $(1+1)$-dimensional transfer matrix geometry.}\label{Fig3}
\end{figure}

In $1+1$ dimensions, the f\/ield theory arising from the ef\/fective
action can be reduced to a~$\mathcal{PT}$-symmetric Hamiltonian acting on class
functions of the gauge group. The ef\/fective action, including the
ef\/fects of heavy quarks, is
\begin{gather*}
S_{\rm ef\/f}=\int d^{2}x\left[\frac{1}{4g^{2}}\left(F_{\mu\nu}^{a}\right)^{2}
-h_{F}\big(e^{\beta\mu}{\rm Tr}_{F}(P)+e^{-\beta\mu}{\rm Tr}_{F}(P^{+})\big)\right],
\end{gather*}
where the gauge f\/ield $A_\mu$ now has two components.
Fig.~\ref{Fig2} shows the Polyakov loop in a $1+1$-dimensional geometry.
It is convenient to work in a gauge where $A_{1}=0$; this is turn
implies that  $A_{2}$ depends only on $x_{1}.$ After integration
over $x_{2}$, we are left with a Lagrangian
\begin{gather*}
L=\frac{\beta}{2g^{2}}\left(\frac{d{A}_{2}^{a}}{dx_{1}}\right)^{2}
-h_{F}\beta\left[e^{\beta\mu}{\rm Tr}_{F}(P)+e^{-\beta\mu}{\rm Tr}_{F}(P^{+})\right],
\end{gather*}
 which we regard as the Lagrangian for a system evolving as a function
of a time coordinate $x_{1}$. This represents a change from a Euclidean
time point of view to a transfer matrix geometry,
as shown in Fig.~\ref{Fig3}.
In this geometry,
the Polyakov loop represents the insertion of an electric f\/lux line
in a box with periodic boundary conditions, and the
free energy density is obtained from the lowest-lying eigenvalue of
the transfer matrix.

The physical states of the system are gauge-invariant, meaning that
they are class functions of $P$: $\Psi\left[P\right]=\Psi\left[gPg^{+}\right]$.
The group characters form an orthonormal basis on the physical Hilbert
space: $\Psi\left[P\right]=\sum_{R}a_{R}{\rm Tr}_{R}\left(P\right)$.
The Hamiltonian $H$, obtained from $L$, acts on the physical states
as \begin{gather*}
H=\frac{g^{2}\beta}{2}C_{2}-h_{F}\beta\left[e^{\beta\mu}{\rm Tr}_{F}(P)+e^{-\beta\mu}{\rm Tr}_{F}(P^{+})\right],
\end{gather*}
where $C_{2}$ is the quadratic Casimir operator for the gauge group,
the Laplace--Beltrami operator on the group manifold.
We have thus reduced the problem of heavy quarks at f\/inite density
in $1+1$ dimensions to one of quantum mechanics on the gauge group.
Unfortunately, the Hamiltonian $H$ is not Hermitian when $\mu\ne0$,
and thus cannot be relied upon to have real eigenvalues. This is a
direct manifestation of the sign problem.

Although the Hamiltonian $H$ is not Hermitian when $\mu \neq 0$, it is $\mathcal{PT}$-symmetric
under the transformations
\begin{gather*}
\mathcal{P}: \ x_{2}\rightarrow-x_{2}, \quad A_{2}\rightarrow-A_{2},\qquad
\mathcal{T}: \ i\rightarrow-i,
\end{gather*}
 which should be regarded as parity and time-ref\/lection in the transfer
matrix geometry.
Together these lead to
\begin{gather*}
\mathcal{PT}:\ P\rightarrow P,
\end{gather*}
which leaves the Hamiltonian invariant.
If this $\mathcal{PT}$ symmetry is unbroken, the eigenvalues
of the Hamiltonian will be real, and there is no sign problem. The
$\mathcal{PT}$ symmetry remains even in the high-density limit where the quark
mass $M$ and chemical potential $\mu$ are taken to inf\/inity in such
a way that antiparticles are suppressed and $P^{+}$ does not appear
in~$H$.

The simplest non-trivial gauge group is ${\rm SU}(3)$, because the cases
of ${\rm U}(1)$ and ${\rm SU}(2)$ are atypical. For the gauge group ${\rm U}(1)$,
the Hamiltonian $H$ may be written as
\begin{gather*}
H=-\frac{e^{2}\beta}{2}\frac{d^{2}}{d\theta^{2}}-h_{F}\beta\big(e^{\beta\mu+i\theta}
+e^{-\beta\mu-i\theta}\big),
\end{gather*}
but a simple change of variable $\theta\rightarrow\theta+i\beta\mu$
eliminates $\mu$:
\begin{gather*}
H=-\frac{e^{2}\beta}{2}\frac{d^{2}}{d\theta^{2}}-h_{F}\beta\big(e^{+i\theta}+e^{-i\theta}\big).
\end{gather*}
This is very similar to the case of the
two-dimensional $\mathcal{PT}$-symmetric sine-Gordon model consi\-de\-red
in~\cite{Bender:2005hf}. In the case of ${\rm SU}(2)$,
all the irreducible representations are real, and the Hamiltonian
is Hermitian:\begin{gather*}
H_{{\rm SU}(2)}=\frac{g^{2}\beta}{2}C_{2}-2h_{F}\cosh\left(\beta\mu\right)\chi_{j=1/2}(P).
\end{gather*}
 This reality feature of ${\rm SU}(2)$ gauge theories at f\/inite density
holds in general, and has been exploited in lattice simulations
with $\mu\ne0$ \cite{Hands:1999md,Kogut:2001na}.

Thus $N=3$ is the f\/irst non-trivial case for ${\rm SU}(N)$ gauge groups.
We have calculated the lowest eigenvalues of $H$ using f\/inite dimensional
approximants. It is convenient to work in the group character basis.
The Casimir operator $C_{2}$ is diagonal in this basis, and characters
act as raising and lowering operators.
For example, in the $4\times4$
subspace spanned by the $1$, $3$, $\bar{3}$, and~$8$ representations
of ${\rm SU}(3)$, the Hamiltonian takes the form
\begin{gather*}
\left(\begin{array}{llll}
0 & e^{-\beta\mu}h_{F}\beta & e^{\beta\mu}h_{F}\beta & 0\\
e^{\beta\mu}h_{F}\beta & \frac{4}{3}\cdot\frac{g^{2}\beta}{2} & e^{-\beta\mu}h_{F}\beta & e^{\beta\mu}h_{F}\beta\\
e^{-\beta\mu}h_{F}\beta & e^{\beta\mu}h_{F}\beta & \frac{4}{3}\cdot\frac{g^{2}\beta}{2} & e^{-\beta\mu}h_{F}\beta\\
0 & e^{-\beta\mu}h_{F}\beta & e^{\beta\mu}h_{F}\beta & 3\cdot\frac{g^{2}\beta}{2}\end{array}\right).
\end{gather*}
If $h_{F}$ is set to zero, we see that the eigenvalues are proportional
to Casimir invariants $0$, $4/3$, $4/3$, and $3$ for the $1$,
$3$, $\bar{3}$, and $8$ representations of ${\rm SU}(3)$.
We have therefore removed an overall factor of $g^{2}\beta/2$,
so the overall strength of the potential term is controlled
by the dimensionless parameter $2h_{F}/g^{2}$. The resulting dimensionless
energy eigenvalues are thus normalized to give the quadratic Casimir operator
when $2h_{F}/g^{2}=0$.
The lowest eigenvalues have been calculated numerically using a basis
of dimension nine or larger, with the stability of the lowest eigenvalues
checked by changing the basis size.

\begin{figure}[t]
\centerline{\includegraphics[width=8.5cm]{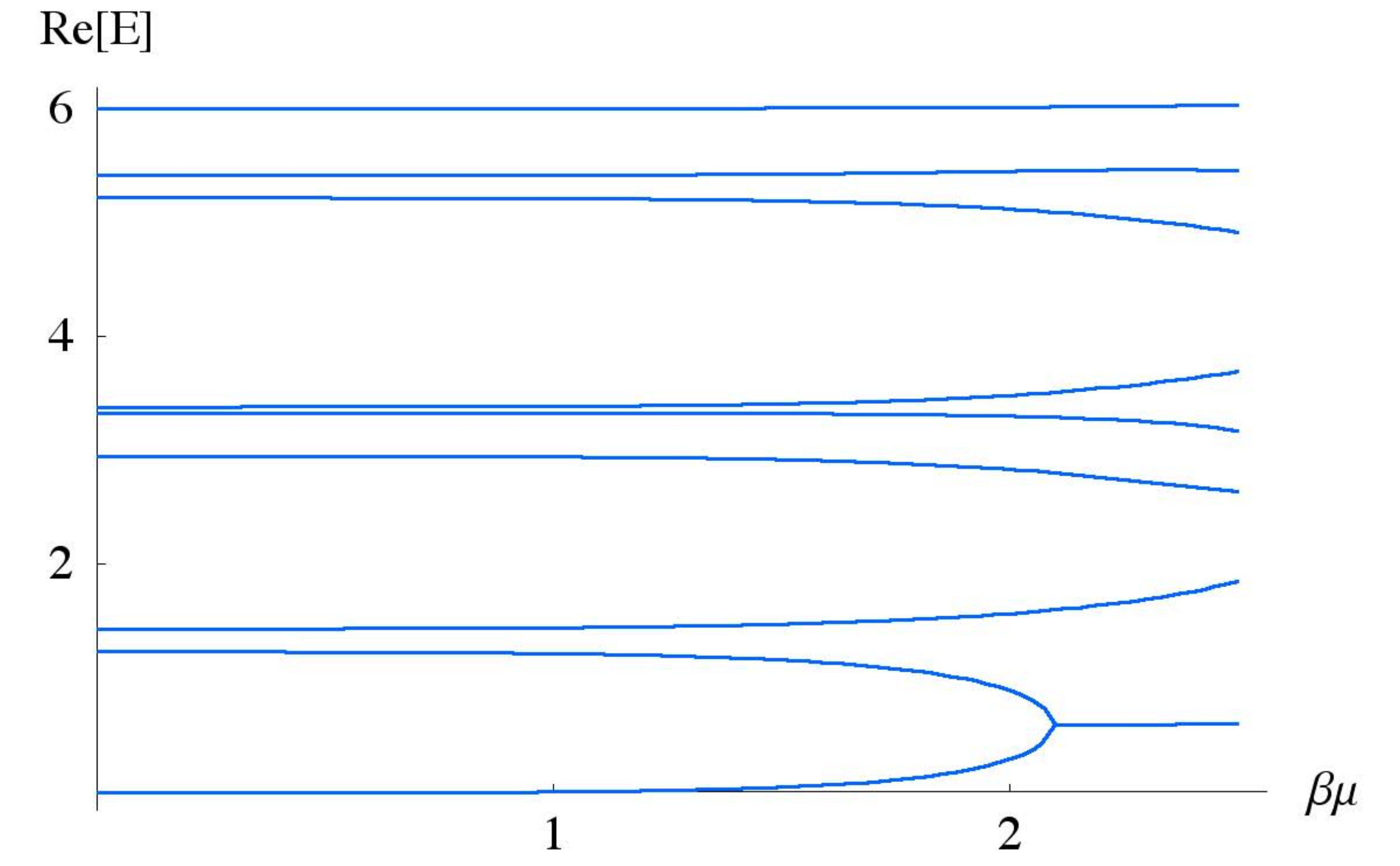}}

\caption{Spectrum for ${2h_{F}}/{g^{2}}=0.1$.}\label{Fig4}

\end{figure}

\begin{figure}[t]
\centerline{\includegraphics[width=8.5cm]{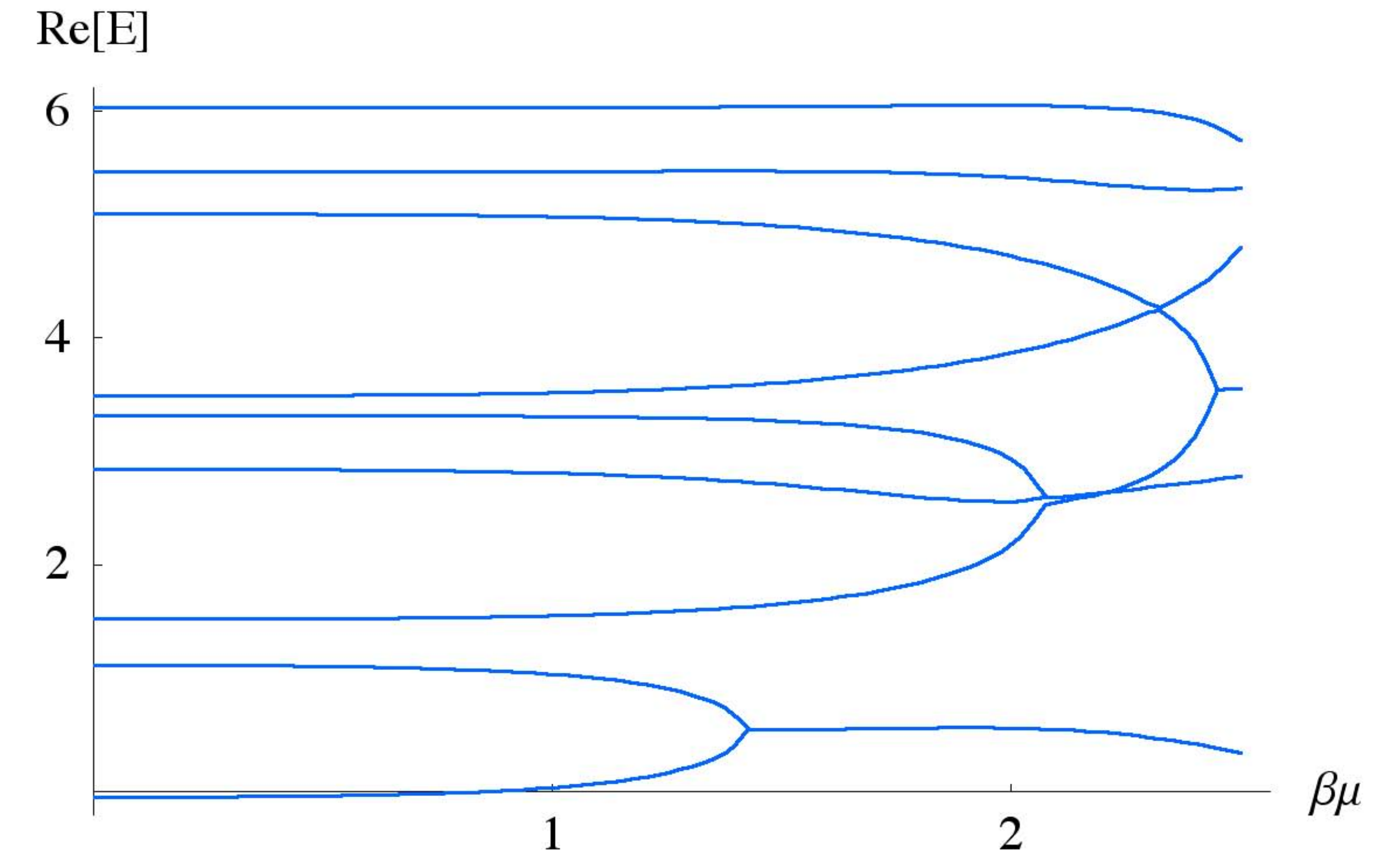}}

\caption{Spectrum for ${2h_{F}}/{g^{2}}=0.2$.}\label{Fig5}

\end{figure}

\begin{figure}[t]
\centerline{\includegraphics[width=8.5cm]{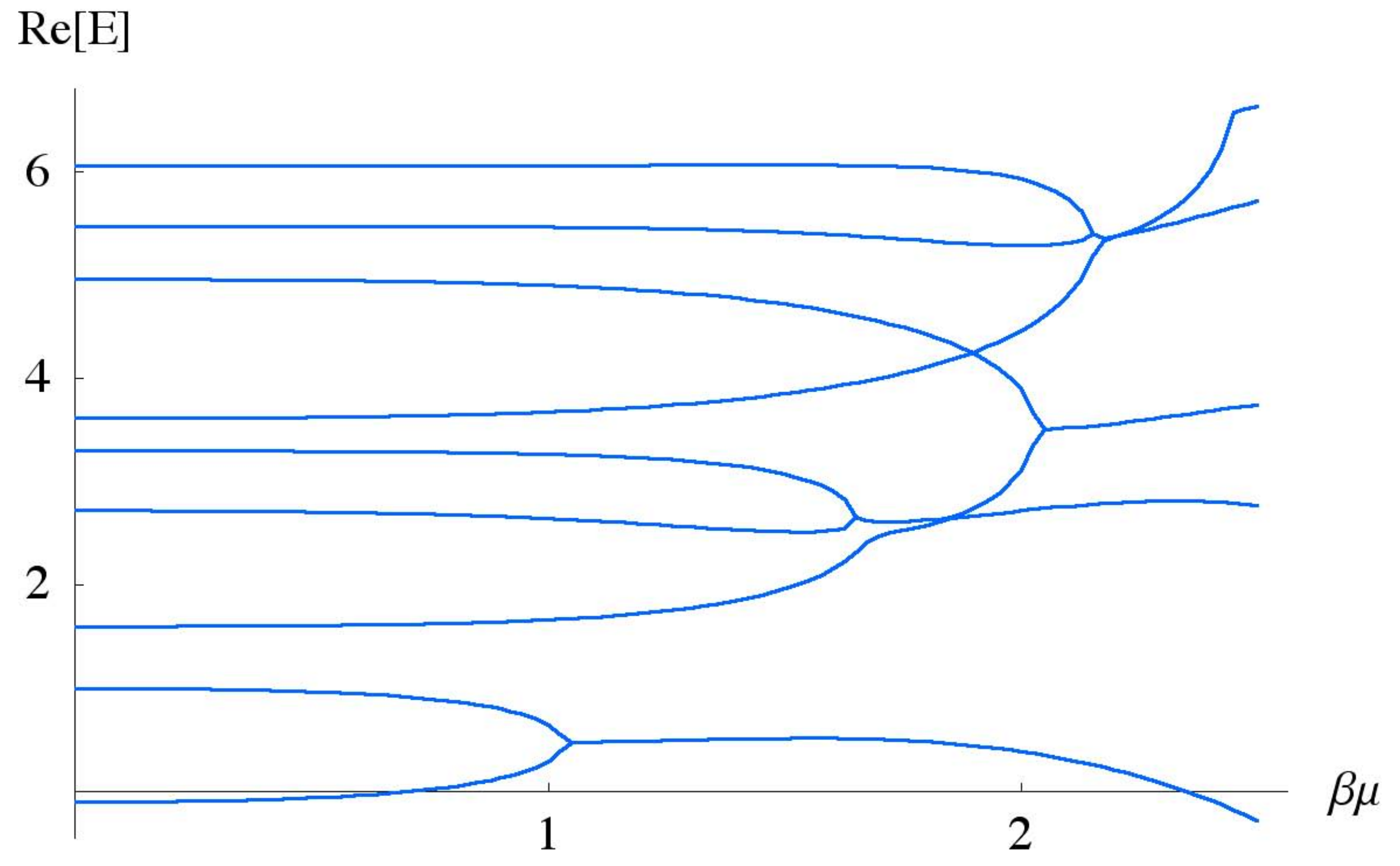}}

\caption{Spectrum for ${2h_{F}}/{g^{2}}=0.3$. }\label{Fig6}

\end{figure}

\begin{figure}[th!]
\centerline{\includegraphics[width=8.5cm]{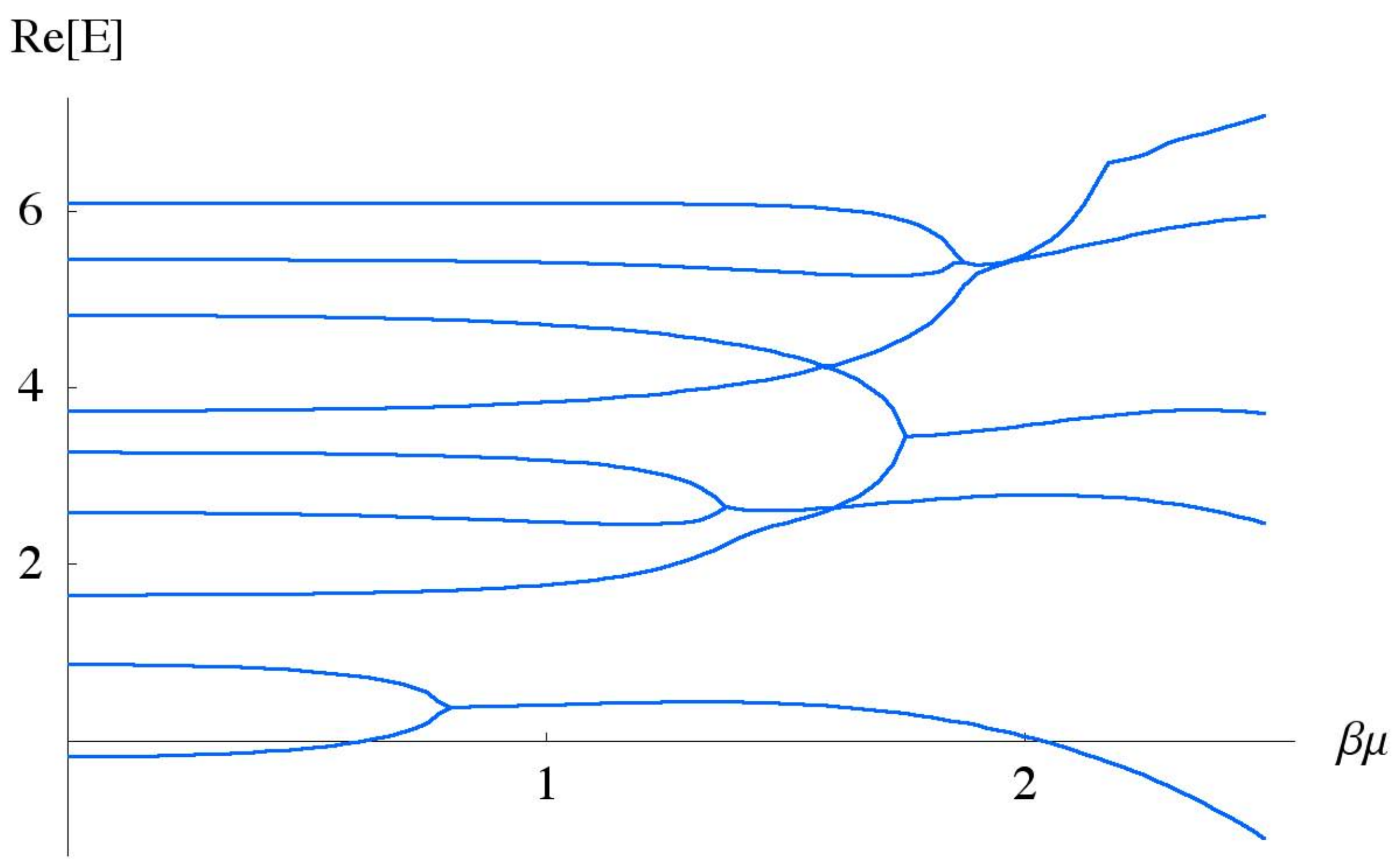}}

\caption{Spectrum for ${2h_{F}}/{g^{2}}=0.4$.}\label{Fig7}

\end{figure}

\begin{figure}[th!]
\centerline{\includegraphics[width=8.5cm]{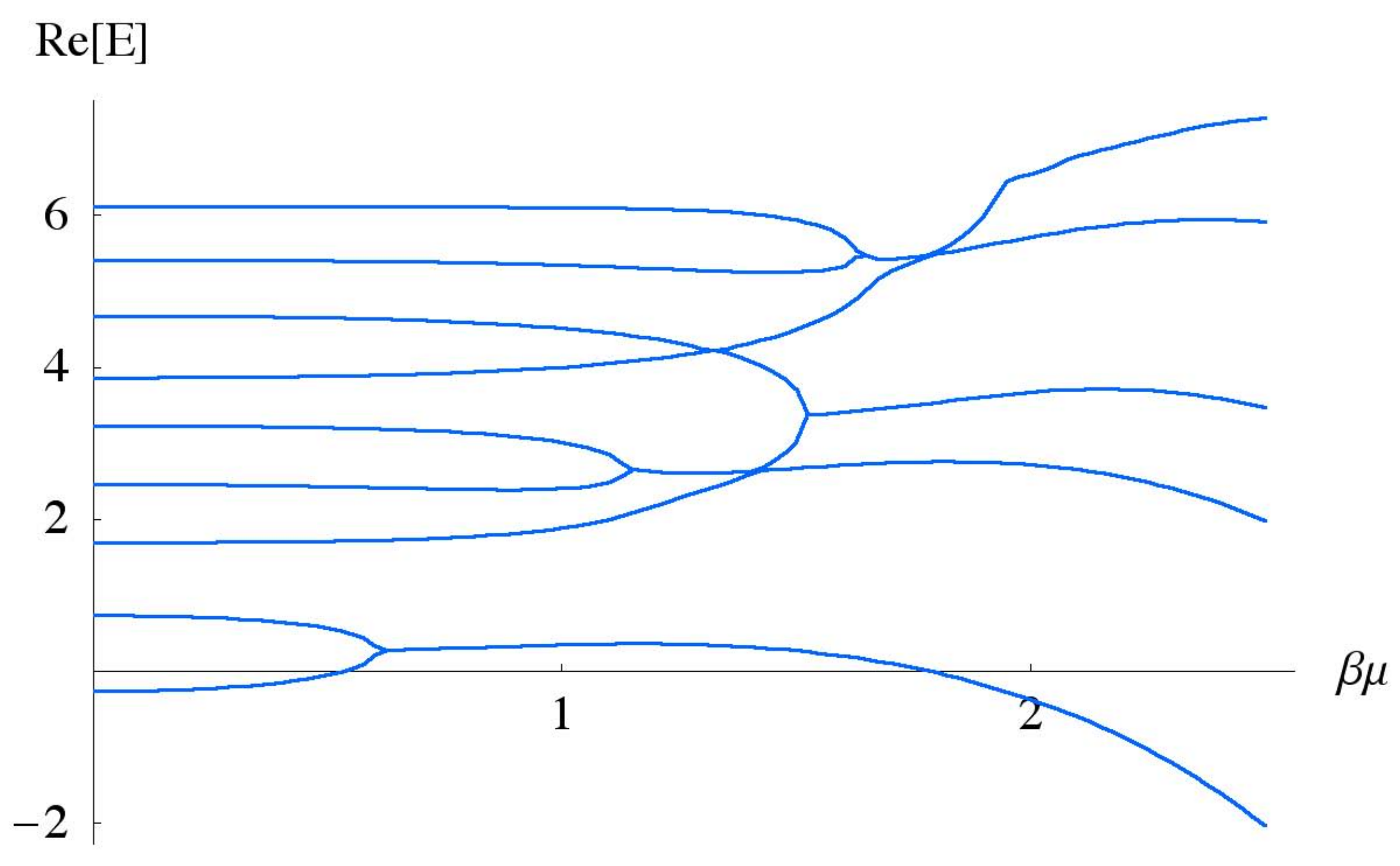}}

\caption{Spectrum for ${2h_{F}}/{g^{2}}=0.5$.}\label{Fig8}

\end{figure}

\begin{figure}[th!]
\centerline{\includegraphics[width=8.5cm]{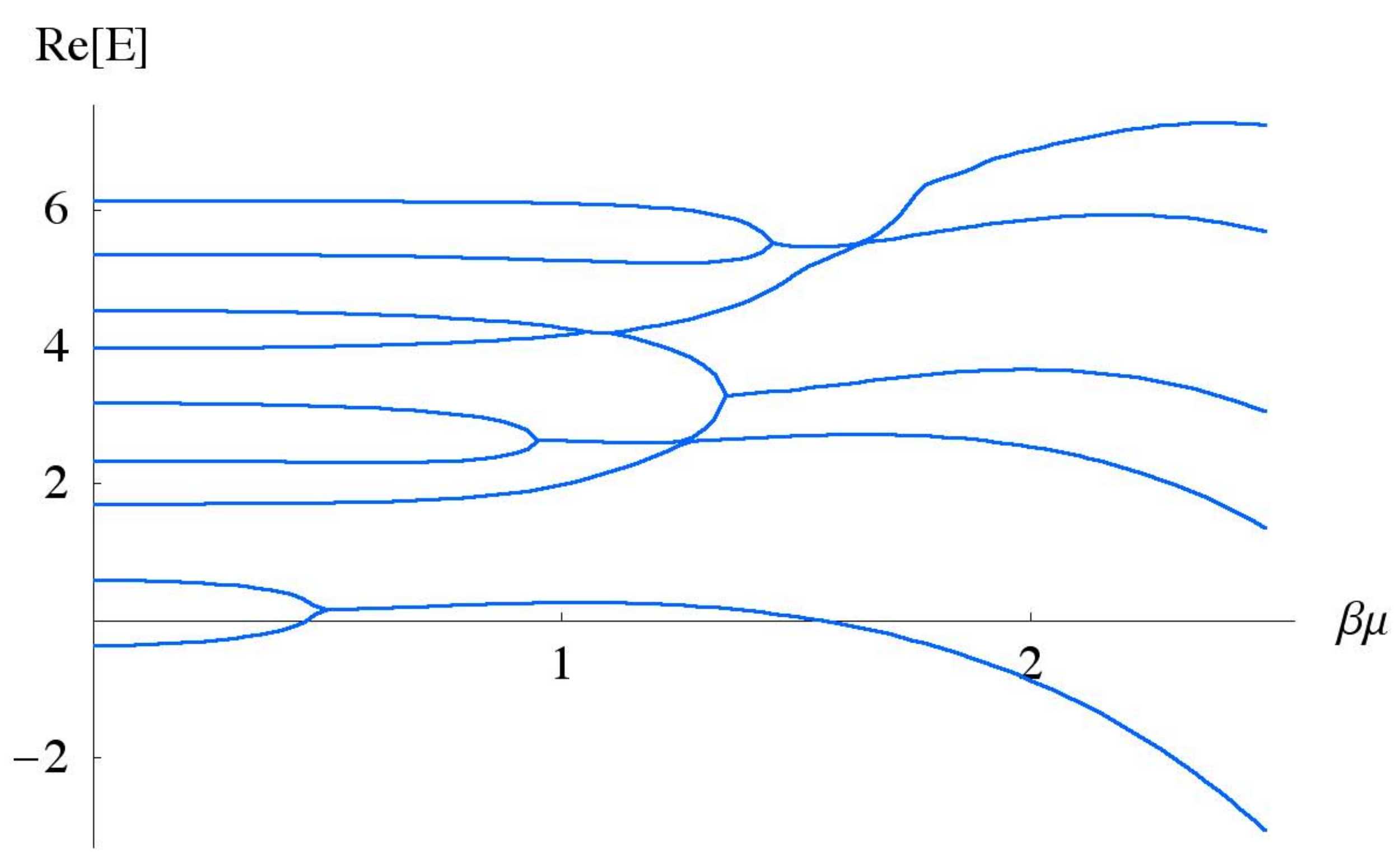}}

\caption{Spectrum for ${2h_{F}}/{g^{2}}=0.6$.}\label{Fig9}

\end{figure}

\begin{figure}[th!]
\centerline{\includegraphics[width=8.5cm]{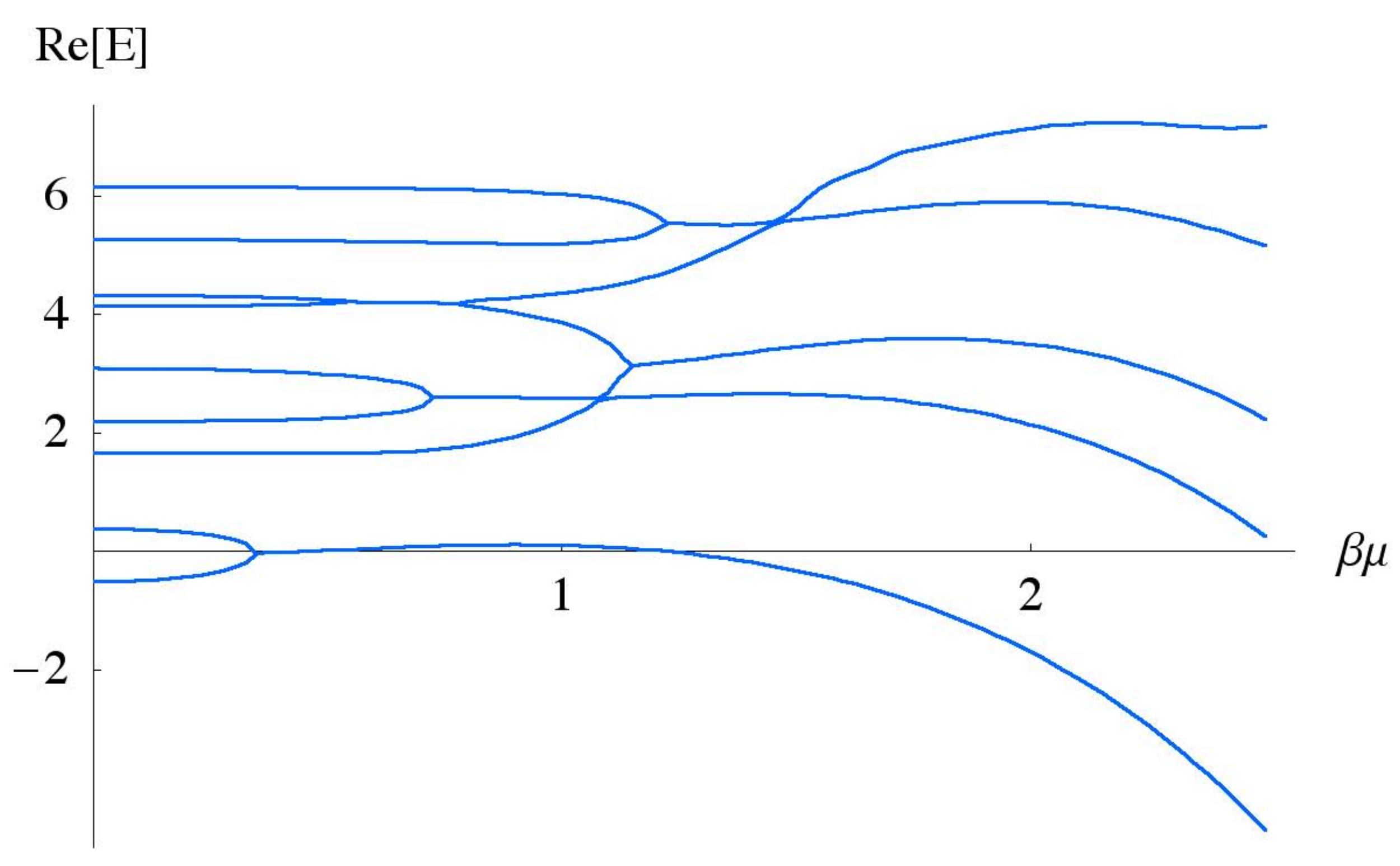}}

\caption{Spectrum for ${2h_{F}}/{g^{2}}=0.75$.}\label{Fig10}

\end{figure}

When $\mu = 0$, the Hamiltonian is Hermitian and all eigenvalues
are guaranteed to be real.
As we see in Fig.~\ref{Fig4}, for $2h_{F}/g^{2}\ll1$ and $\mu=0$, the
eigenvalues are close to the quadratic Casimir values. Even when $\mu=0$,
the ef\/fect of the quarks is to split the degeneracies found in the
pure gauge theory. Thus, linear combinations of the $3$ and $\bar{3}$
representations show a splitting. As $\mu$ increases from zero, the eigenvalues
remain real. Eventually, the two lowest
energy eigenvalues approach one another, forming a complex conjugate
pair indicating the breaking of $\mathcal{PT}$ symmetry.
As shown in Figs.~\ref{Fig4}--\ref{Fig10}, increasing
the parameter $2h_{F}/g^{2}$ decreases the value of
$\beta\mu$ at which the coalescence of the
two lowest eigenvalues occurs.
Unlike the $\left(-ix\right)^{p}$
models where the lowest energy eigenvalues are the last to become
complex as the parameter $p$ decreases, here $\mathcal{PT}$ symmetry breaking
appears to occur in the lowest energies f\/irst.
Figs.~\ref{Fig4}--\ref{Fig10} show that the eigenvalues above
the ground state have a similar complicated behavior as a function of $\beta\mu$.
However, the free energy density
in the limit of inf\/inite spatial dimension
only depends on the ground state energy
in the transfer matrix geometry.
It is not yet clear what, if any,
is the physical meaning of the breaking of $\mathcal{PT}$ symmetry in this
context. Nevertheless, it is clear that there is a range of values
for $\beta\mu$ for which the sign problem is avoided, due to $\mathcal{PT}$
symmetry.

The partition function associated with the ef\/fective Lagrangian
$L$ can also be interpretated
as a classical statistical mechanical system. For simplicity, consider
the case of the ${\rm U}(1)$ gauge group. The partition function can be
expanded as a power series in $\beta h_{F}$, and the path integral
over $A_{2}$ performed order by order. The result in the ${\rm U}(1)$
case is\begin{gather*}
Z=\sum_{n=0}^{\infty}\frac{\left(\beta h_{F}\right)^{2n}}{\left(n!\right)^{2}}\int dy_{1}\cdots dy_{n}dz_{1}\cdots dz_{n}\\
\phantom{Z=}{}\times \exp\left\{ \frac{e^{2}\beta}{2}\sum_{j\ne k}\left[G\left(y_{j}-z_{k}\right)-G\left(y_{j}-y_{k}\right)-G\left(z_{j}-z_{k}\right)\right]\right\},
\end{gather*}
 where $G(x)$ is the one-dimensional Green's function $G(x)=-\left(1/2\right)\left|x\right|$.
This is an example of the familiar equivalence between f\/ield theories
of sine-Gordon type and the classical Coulomb gas.
Although f\/irst derived in one dimension~\cite{Edwards:1962zz}, the equivalence
holds in all dimensions. More complicated gauge groups result in a similar,
but more complicated expansion with a non-trivial dependence
on $\mu$. As will be discussed below, generalized non-Hermitian sine-Gordon
models are also relevant for the study of quark conf\/inement.

\section[Confinement]{Conf\/inement}

It is remarkable that there are two classes of $Z(N)$-invariant
systems that are conf\/ining at arbitrarily high temperatures, evading
the transition to the deconf\/ined phase found in the pure gauge theory.
Both classes obtain
a high-temperature conf\/ined phase from a pure gauge theo\-ry
by the addition of fermions in the adjoint representation of ${\rm SU}(N)$,
with the non-standard choice of periodic boundary conditions
for the fermions in the
timelike direction.
One class consists of $\mathcal{N}=1$ supersymmetric gauge
theories \cite{Davies:1999uw,Davies:2000nw},
where the periodic boundary conditions on the gauginos
is necessary to preserve supersymmetry.
The perturbative contribution to the ef\/fective potential for
the Polyakov loop is identically zero, because the gauge f\/ield
contribution is cancelled exactly by the gaugino contribution.
However, the non-perturbative contribution to the ef\/fective potential
can be calculated exactly, and leads to a single, $Z(N)$-invariant,
conf\/ining phase.
These ideas have recently been extended to
a second class of models where
the ef\/fect of adjoint fermions dominates
the contribution of the gauge f\/ields
\cite{Myers:2007vc,Unsal:2007vu,Unsal:2008ch}.
If the number of adjoint fermion f\/lavors $N_{f}$ is not too large,
these systems are asymptotically free
at high temperature, and therefore
the ef\/fective potential for $P$ is calculable using perturbation theory.
The system
will lie in the conf\/ining phase
if the fermion mass $m$ is suf\/f\/iciently light and $N_{f} > 1/2$.
In this case, electric string tensions can be calculated
perturbatively from the ef\/fective potential.
In~both this case and the supersymmetric case, magnetic string tensions
arise semiclassically from non-Abelian magnetic monopoles. This
provides a realization of one of
the oldest ideas about the origin of conf\/inement.
Moreover, lattice simulations \cite{Myers:2007vc} indicate
that this high-temperature conf\/ining region is smoothly connected
to the low-temperature conf\/ining phase of the pure gauge theory
as the temperature is lowered and the fermion mass is increased.

Up to a point, both classes of models can be treated similarly, but we will
largely focus on the second, non-supersymmetric case.
The one-loop ef\/fective potential for a boson in a~representa\-tion~$R$
with spin degeneracy $s$ moving in a Polyakov loop background $P$ at
non-zero temperature and density is given by \cite{Gross:1980br,Weiss:1980rj}
\begin{gather*}
V_{b}=sT\int\frac{d^{d}k}{\left(2\pi\right)^{d}}{\rm Tr}_{R}\left[\ln\big(1-Pe^{\beta\mu-\beta\omega_{k}}\big)+
\ln\big(1-P^{+}e^{-\beta\mu-\beta\omega_{k}}\big)\right].
\end{gather*}
Periodic boundary conditions are assumed. With standard boundary conditions
(periodic for bosons, antiperiodic for fermions), 1-loop ef\/fects always
favor the deconf\/ined phase. For the case of pure gauge theories, the
one-loop ef\/fective potential can be written in the form
\begin{gather*}
V_{\rm gauge}\left(P,\beta,m,N_{f}\right)=\frac{-2}{\pi^{2}\beta^{4}}\sum_{n=1}^{\infty}\frac{{\rm Tr}_{A}P^{n}}{n^{4}}.
\end{gather*}
This series is minimized, term by term if $P\in Z(N)$, so $Z(N)$
symmetry is spontaneously broken at high temperature. The same result
is obtained for any bosonic f\/ield with periodic boundary conditions
or for fermions with antiperiodic boundary conditions.

The addition of fermions with periodic boundary conditions can restore
the broken $Z(N)$ symmetry. Consider the case of $N_{f}$ f\/lavors
of Dirac fermions in the adjoint representation of ${\rm SU}(N)$. Periodic
boundary conditions in the timelike direction imply that the generating
function of the ensemble, i.e., the partition function, is
given by\begin{gather*}
Z={\rm Tr}\big[\left(-1\right)^{F}e^{-\beta H}\big],
\end{gather*}
where $F$ is the fermion number. This ensemble, familiar from supersymmetry,
can be obtained from an ensemble at chemical potential $\mu$ by the
replacement $\beta\mu\rightarrow i\pi$. In perturbation theory, this
shifts the Matsubara frequencies from $\beta\omega_{n}=\left(2n+1\right)\pi$
to $\beta\omega_{n}=2n\pi$. The one loop ef\/fective potential is like
that of a bosonic f\/ield, but with an overall negative sign due to
fermi statistics \cite{Meisinger:2001fi}. The sum of the ef\/fective potential for the
fermions plus that of the gauge bosons gives
\begin{gather*}
V_{\text{1-loop}}\left(P,\beta,m,N_{f}\right)=\frac{1}{\pi^{2}\beta^{4}}
\sum_{n=1}^{\infty}\frac{{\rm Tr}_{A}P^{n}}{n^{2}}\left[2N_{f}\beta^{2}m^{2}K_{2}\left(n\beta m\right)-\frac{2}{n^{2}}\right].
\end{gather*}
Note that the f\/irst term in brackets, due to the fermions, is positive
for every value of $n$, while the second term, due to the gauge bosons,
is negative.

The largest contribution to the ef\/fective potential at high temperatures
is typically from the $n=1$ term, which can be written simply as
\begin{gather*}
\frac{1}{\pi^{2}\beta^{4}}\big[2N_{f}\beta^{2}m^{2}K_{2}\left(\beta m\right)-2\big]\big[\left|{\rm Tr}_{F}P\right|^{2}-1\big],
\end{gather*}
 where the overall sign depends only on $N_{f}$ and $\beta m$. If
$N_{f} >1/2$ and $\beta m$ is suf\/f\/iciently small, this term will
favor ${\rm Tr}_{F}P=0$. On the other hand, if $\beta m$ is suf\/f\/iciently
large, a value of~$P$ from the center, $Z(N)$, is preferred. Note
that an $\mathcal{N}=1$ super Yang--Mills theory would correspond to $N_{f}=1/2$ and
$m=0$, giving a vanishing perturbative contribution for all~$n$~\mbox{\cite{Davies:1999uw,Davies:2000nw}}. In this case,
it is necessary to calculate the non-perturbative contribution
to the ef\/fective potential.
This suggests that it should be possible to obtain a $Z(N)$ symmetric,
conf\/ining phase at high temperatures using adjoint fermions with periodic
boundary conditions or some equivalent deformation of the theory.

This possibility has been conf\/irmed in ${\rm SU}(3)$, where both lattice
simulations and perturbative calculations have been used to show that
a gauge theory action with an extra term of the form $\int d^{4}x\, a_{1}{\rm Tr}_{A}P$
is conf\/ining for suf\/f\/iciently large $a_{1}$ at arbitrarily high temperatures~\cite{Myers:2007vc}.
This simple, one-term deformation is suf\/f\/icient for ${\rm SU}(2)$ and
${\rm SU}(3)$. However, in the general case, a~deformation with at least
$\left[\frac{N}{2}\right]$ terms is needed to assure conf\/inement
for representations of all possible non-zero $k$-alities. Thus the
minimal deformation necessary is of the form
\begin{gather*}
\sum_{k=1}^{\left[\frac{N}{2}\right]}a_{k}{\rm Tr}_{A}P^{k},
\end{gather*}
which is analyzed in detail in \cite{Myers:2008ey}. If all the coef\/f\/icients
$a_{k}$ are suf\/f\/iciently large and positive, the free energy density
\begin{gather*}
V_{\text{1-loop}}\left(P,\beta,m,N_{f}\right)=
\frac{-2}{\pi^{2}\beta^{4}}\sum_{n=1}^{\infty}\frac{{\rm Tr}_{A}P^{n}}{n^{4}}
+\sum_{k=1}^{\left[\frac{N}{2}\right]}a_{k}{\rm Tr}_{A}P^{k}
\end{gather*}
will be minimized by a unique set of Polyakov loop eigenvalues corresponding
to exact $Z(N)$ symmetry.

The unique set of eigenvalues of $P$ invariant under $Z(N)$
is $\left\{ w,wz,wz^{2},\dots,wz^{N-1}\right\} $, where $z=e^{2\pi i/N}$ is
the generator of $Z(N)$, and $w$ is a phase necessary to ensure
unitarity \cite{Meisinger:2001cq}. A~matrix with these eigenvalues, such as
$P_{0}=w\cdot {\rm diag}\left[1,z,z^{2},\dots,z^{N-1}\right]$,
 is gauge-equivalent to itself after a $Z(N)$ symmetry operation:
$zP_{0}=gP_{0}g^{+}$.
This guarantees that
 ${\rm Tr}_{F}\left[P_{0}^{k}\right]=0$
for any value of~$k$ not divisible by $N$,
indicating conf\/inement for all representations transforming
non-trivially under $Z(N)$.

To f\/ind the conditions under which $P_{0}$ is a global minimum of the ef\/fective potential,
we use the high-temperature expansion for the one-loop free energy
of a particle in an arbitrary background Polyakov loop gauge equivalent
to the matrix $P_{jk}=\delta_{jk}e^{i\phi_{j}}$.
The f\/irst two terms have the form \cite{Meisinger:2001fi}
\begin{gather*}
V_{\text{1-loop}} \approx \sum_{j,k=1}^{N}\left(1-\frac{1}{N}\delta_{jk}\right)
\frac{2\left(2N_{f}-1\right)T^{4}}{\pi^{2}}\left[\frac{\pi^{4}}{90}
-\frac{1}{48\pi^{2}}\left(\phi_{j}-\phi_{k}\right)^{2}\left(\phi_{j}-\phi_{k}-2\pi\right)^{2}\right]\nonumber\\
\phantom{V_{\text{1-loop}} \approx}{}
-\sum_{j,k=1}^{N}(1-\frac{1}{N}\delta_{jk})\frac{N_{f}m^{2}T^{2}}{\pi^{2}}
\left[\frac{\pi^{2}}{6}+\frac{1}{4}\left(\phi_{j}-\phi_{k}\right)\left(\phi_{j}-\phi_{k}-2\pi\right)\right].
\end{gather*}
The $T^{4}$ term dominates for $m/T\ll1$,
and has $P_{0}$ as a minimum provided $N_{f}>1/2$. Even if the adjoint
fermion mass is enhanced by chiral symmetry breaking, as would be
expected in a~conf\/ining phase, it should be of order $gT$ or less,
and the second term in the expansion of $V_{\text{1-loop}}$ can be neglected
at suf\/f\/iciently high temperature. It is interesting to note that
for $N_{f}=1/2$, any $m>0$ will give a perturbative term
that leads to a deconf\/ined phase.

\section{String tension scaling laws}

The timelike string tension $\sigma_{k}^{(t)}$ between $k$ quarks and
$k$ antiquarks can be measured from the behavior of the correlation
function \begin{gather*}
\langle {\rm Tr}_{F}P^{k}\left(\vec{x}\right){\rm Tr}_{F}\left(P^{+}\left(\vec{y}\right)\right)^{k}\rangle
\simeq\exp\left[-\frac{\sigma_{k}^{(t)}}{T}\left|\vec{x}-\vec{y}\right|\right]\end{gather*}
at suf\/f\/iciently large distances. Two widely-considered scaling behaviors
for string tensions are Casimir scaling, characterized by
\begin{gather*}
\sigma_{k}=\sigma_{1}\frac{k\left(N-k\right)}{N-1},
\end{gather*}
 and sine-law scaling, given by\begin{gather*}
\sigma_{k}=\sigma_{1}{\frac{\sin\left[\pi k/N\right]}{\sin\left[\pi/N\right]}}.
\end{gather*}
 For a review, see reference \cite{Greensite:2003bk}.

At non-zero temperatures, time-like and space-like string tensions may be dif\/ferent. Time-like string tensions may be measured by Polyakov loop correlation functions. while spatial string tensions are measured by space-like Wilson loops. For the supersymmetric case, both string tensions obey  sine-law scaling. This is a consequence of the close connection of this class of models with the af\/f\/ine Toda f\/ield theory, which is a $\mathcal{PT}$-symmetric model~\cite{Hollowood:1992by}. We will explore the string tension scaling laws  for the second class of models, with
$N_f > 1/2$, and then return to the af\/f\/ine Toda models and their possible connection to QCD.

Timelike string tensions are calculable perturbatively
in the high-temperature conf\/ining region for $N_f > 1/2$
from small f\/luctuations about
the conf\/ining minimum of the ef\/fective potential~\cite{Meisinger:2004pa}.
The scale is naturally of order $gT$:
\begin{gather*}
\left(\frac{\sigma_{k}^{(t)}}{T^2}\right)^2
 = g^{2}N\frac{N_{f}m^{2}}{\pi^{2}}\sum_{j=0}^{\infty}\big[K_{2}\left((k+jN)\beta m\right)+K_{2}\left((N-k+jN)\beta m\right)\\
\phantom{\left(\frac{\sigma_{k}^{(t)}}{T^2}\right)^2=}{}-2K_{2}\left((j+1)N\beta m\right)\big] -g^{2}N\frac{T^{2}}{3N^{2}}\left[3\csc^{2}\left(\frac{\pi k}{N}\right)-1\right].
\end{gather*}
 These string tensions are continuous functions of $\beta m$. The
$m=0$ limit is simple: \begin{gather*}
\left(\frac{\sigma_{k}^{(t)}}{T}\right)^{2}=\frac{\left(2N_{f}-1\right)g^{2}T^{2}}{3N}\left[3\csc^{2}\left(\frac{\pi k}{N}\right)-1\right]\end{gather*}
and is a good approximation for $\beta m\ll1$. This scaling law is
not at all like either Casimir or sine-law scaling, because the usual
hierarchy $\sigma_{k+1}^{(t)}\ge\sigma_{k}^{(t)}$ is here reversed.
Because we expect on the basis of ${\rm SU}(3)$ simulations that the high-temperature
conf\/ining region is continuously connected to the conventional low-temperature
region, there must be an inversion of the string tension hierarchy
between the two regions for all $N\ge4$.

The conf\/ining minimum $P_{0}$ of the ef\/fective potential breaks ${\rm SU}(N)$
to ${\rm U}(1)^{N-1}$. This remaining unbroken Abelian gauge group naively
seems to preclude spatial conf\/inement, in the sense of area law behavior
for spatial Wilson loops. However, as f\/irst discussed by Polyakov
in the case of an ${\rm SU}(2)$ Higgs model in $2+1$ dimensional gauge
systems, instantons can lead to nonperturbative conf\/inement~\cite{Polyakov:1976fu}.
In the high-temperature conf\/ining region, the dyna\-mics of the magnetic
sector are ef\/fectively three-dimensional due to dimensional reduction.
The Polyakov loop plays a role similar to an adjoint Higgs f\/ield,
with the important dif\/ference that $P$ lies in the gauge group, while
a Higgs f\/ield would lie in the gauge algebra. The standard topological
analysis \cite{Weinberg:1979zt} is therefore slightly altered, and there are $N$ fundamental
monopoles in the f\/inite temperature gauge theory
\cite{Lee:1998vu,Kraan:1998kp,Lee:1998bb,Kraan:1998pm,Kraan:1998sn} with charges
proportional to the af\/f\/ine roots of~${\rm SU}(N)$, given by
$2\pi\alpha_{j}/g$
where
$\alpha_{j}=\hat{e}_{j}-\hat{e}_{j+1}$
for $j=1$ to $N-1$ and
$\alpha_{N}=\hat{e}_{N}-\hat{e}_{1}.$
Monopole ef\/fects will be suppressed by powers of the Boltzmann factor
$\exp\left[-E_{j}/T\right]$ where $E_{j}$ is the energy of a monopole
associated with $\alpha_{j}$.

In the high-temperature conf\/ining region,
monopoles interact with each other through both their long-ranged
magnetic f\/ields, and also via a three-dimensional scalar interaction, mediated by $A_{4}$.
The scalar interaction is short-ranged,
falling of\/f with a mass of order  $gT$. The long-range properties of the magnetic sector
may be represented in a simple form by a generalized sine-Gordon model
which generates the grand canonical ensemble for the monopole/anti-monopole
gas \cite{Unsal:2008ch}. The action for this model represents the Abelian dual form
of the magnetic sector of the ${\rm U}(1)^{N-1}$ gauge theory. It is given
by
\begin{gather*}
S_{\rm mag}=\int d^{3}x\left[\frac{T}{2}\left(\partial\rho\right)^{2}-2\xi
\sum_{j=1}^{N}\cos\left(\frac{2\pi}{g}\alpha_{j}\cdot\rho\right)\right],
\end{gather*}
 where $\rho$ is the scalar f\/ield dual to the ${\rm U}(1)^{N-1}$ magnetic
f\/ield. The monopole fugacity $\xi$ is given by $\exp\left[-E_{j}/T\right]$
times functional determinantal factors~\cite{Zarembo:1995am}.

This Lagrangian is a generalization of the one considered by Polyakov
for ${\rm SU}(2)$, and the analysis of magnetic conf\/inement follows along
the same lines \cite{Polyakov:1976fu}. The Lagrangian has $N$ degenerate inequivalent minima
$\rho_{0k}=g\mu_{k}$ where the $\mu_{k}$'s are the simple fundamental
weights, satisfying $\alpha_{j}\cdot\mu_{k}=\delta_{jk}$. Note that
$e^{2\pi i\mu_{k}}=z^{k}$. A spatial Wilson loop
\begin{gather*}
W\left[\mathcal{C}\right]=\mathcal{P}\exp\left[i\oint_{\mathcal{C}}dx_{j}\cdot A_{j}\right]
\end{gather*}
in the $x$-$y$ plane
introduces a discontinuity in the z direction in the f\/ield dual to
$B$.
Moving this discontinuity out to spatial inf\/inity, the string
tension of the spatial Wilson loop is the interfacial energy of a
one-dimensional kink interpolating between the vacua~$\rho_{0k}$.
The calculation is similar to that of the 't Hooft loop in the deconf\/ined phase,
where the kinks interpolate between the~$N$ dif\/ferent solutions associated
with the spontaneous breaking of~${\rm SU}(N)$.
The main technical dif\/f\/iculty lies in f\/inding the correct kink solutions.
A straight line ansatz through
the Lie algebra~\cite{Giovannangeli:2001bh} using 
$\rho(z)=g\mu_{k}q(z)$ 
 gives \begin{gather*}
\sigma_{k}^{(s)}=\frac{8}{\pi}\left[\frac{g^{2}T\xi}{N}k\left(N-k\right)\right]^{1/2}.
\end{gather*}
This result is exact for $N=2$ or $3$, but may be only an upper bound for
$N>3$. The square-root-Casimir scaling behavior obtained
dif\/fers somewhat from
both Casimir and sine-law scaling, and
is inconsistent with lattice simulation results
for $N\ge 4$ pure gauge theories~\cite{Lucini:2004my}.
It is not completely surprising that the addition
of additional particles, in this case adjoint fermions,
might change string tension scalings laws.

Nevertheless, the behavior of spatial string tensions in the high-temperature
conf\/ining region is similar to what we think might occur in the
low-temperature conf\/ining phase of a pure gauge theory.
Therefore it is interesting to ask whether there are classes of
models, corresponding to dif\/ferent monopole gases and having dif\/ferent
string tension scaling laws. The model just discussed is an af\/f\/ine
sine-Gordon model with a sum over the af\/f\/ine simple roots,
both positive and negative.
If we sum instead over the af\/f\/ine positive roots, we have the
af\/f\/ine Toda model, a~non-Hermitian but $\mathcal{PT}$-symmetric, model. The action is
\begin{gather*}
S_{\rm Toda}=\int d^{3}x\left[\frac{T}{2}\left(\partial\rho\right)^{2}-\xi\sum_{j=1}^{N}\exp\left(i\frac{2\pi}{g}
\alpha_{j}\cdot\rho\right)\right].
\end{gather*}
This is an ef\/fective f\/ield theory for a gas of monopoles, but no anti-monopoles.
As shown by Hollowood \cite{Hollowood:1992by}, the kink
solutions of this model have a sine-law mass spectrum:
\begin{gather*}
\sigma_{k}^{(s)}=\frac{2N}{\pi}\left[g^{2}T\xi\right]^{1/2}\sin\left(\frac{\pi k}{N}\right).
\end{gather*}
Diakonov and Petrov have shown that this ef\/fective theory may
be plausibly obtained from ${\rm SU}(N)$ gauge theories at f\/inite temperature
if anti-monopoles are excluded from the ensemble of f\/ield conf\/igurations
considered in the path integral~\cite{Diakonov:2007nv}.
We can also consider a sine-Gordon model with a sum over all roots.
In this case, the string tension exhibits Casimir scaling:
\begin{gather*}
\sigma_{k}^{(s)}=\frac{8}{\pi}\left[\frac{g^{2}T\xi}{N}\right]^{1/2}k\left(N-k\right).
\end{gather*}
The similarity of these three sets of results, and the closeness of lattice simulation
results to both Casimir and sine-law scaling, suggest the possibility of a crossover
from one string tension scaling law to another as the character of the monopole gas
changes. The details of how this might happen, however, are not clear.

\section{Conclusions}

The common thread connecting $\mathcal{PT}$ symmetry to
applications to QCD at non-zero temperature and density
is the use of generalized sine-Gordon models
to represent the statistical mechanics of objects
carrying non-Abelian electric and magnetic charge.
Heavy quarks at non-zero density give rise
to a $\mathcal{PT}$-symmetric ef\/fective action.
This is turn gives us a new way of looking at the sign problem.
In the problem of quark conf\/inement,
there are models
where monopole gases are responsible
for conf\/inement.
Depending on the specif\/ic model,
the ef\/fective action for the monopole gas
may be Hermitian, or non-Hermitian but $\mathcal{PT}$-symmetric.
In addition to the connection of $\mathcal{PT}$ symmetry to QCD,
these models also suggest interesting possibilities for
$\mathcal{PT}$ symmetry more generally in statistical physics.

\pdfbookmark[1]{References}{ref}
\LastPageEnding

\end{document}